\documentstyle[editedvolume,numreferences,psfig]{crckapb}

\begin{opening}
\title{Quantum Phase Transitions}
\author{D.Belitz}
\institute{Department of Physics and Materials Science Institute\\
 University of Oregon\\
 Eugene, OR 97403}
\author{T.R.Kirkpatrick}
\institute{Institute for Physical Science and Technology\\
 and Department of Physics\\
 University of Maryland, College Park, MD 20742}
\end{opening}
\begin{document}
\begin{abstract} These are notes for lectures delivered at
the NATO ASI on Dynamics in Leiden, The Netherlands, in July
1998. The main concepts relating to quantum phase
transitions are explained, using the paramagnet-to-ferromagnet
transition of itinerant electrons as the primary example. 
Some aspects of metal-insulator transitions are also briefly
discussed. The
exposition is strictly pedagogical in nature, with no ambitions
with respect to completeness or going into technical details.
The goal of the lectures is to provide a bridge between textbooks on
classical critical phenomena and the current literature on
quantum phase transitions. Some familiarity with the concepts
of classical phase transitions is helpful, but not absolutely
necessary.

\vskip 25mm

\noindent
Contribution to:\quad  {\it Dynamics: Models and Kinetic Methods for Non-}
\par\noindent
\hskip 51mm {\it equilibrium Many-Body Systems}
\par\noindent
\hskip 32mm John Karkheck (editor)
\par\noindent
\hskip 32mm To be published by Kluwer academic publishers b.v.
\bigskip
\par\noindent
\hskip 32mm Lectures delivered by D. Belitz

\end{abstract}

\section{Introduction to Continuous Phase Transitions}
\label{sec:I}

In these lectures we discuss so-called quantum phase transitions
or zero-temperature phase transitions. Our motivation is the fact that
these transitions provide a field where the long-range correlations
in quantum systems that were introduced in the 
preceding lectures \cite{TRK's_lectures} have
some spectacular consequences. Since 
quantum phase transitions have many features in common with ordinary 
classical or thermal
phase transitions, we will first review the basic concepts of both,
and then define and discuss the particular features of quantum
transitions, before
turning to some specific examples. We would like to warn the reader
that it is not possible to do this subject justice within the time
and space constraints of these lectures. Our aim can therefore only
be to give some examples that highlight a few of the interesting
features of the field. Thorough treatments of the
phenomenology and theory of classical
phase transitions can be found in Refs.\ \cite{ClassicalPhaseTransitions},
and aspects of quantum phase transitions are reviewed in 
Refs.\ \cite{usMarch,QuantumPhaseTransitions}.

\subsection{Review of basic concepts}
\label{subsec:I.1}

\subsubsection{Continuous phase transitions}
\label{subsubsec:I.1.A}

Continuous phase transitions
\footnote[1]{We will deal only with
continuous transitions, and will simply refer to them as `phase transitions' 
or `critical points'.}
are characterized by the system under
consideration undergoing a transition from a symmetric or disordered
state, which incorporates some symmetry of the Hamiltonian, to a
broken-sym\-me\-try or ordered state, which does not have that symmtry,
although the Hamiltonian still possesses it. A good example is given
by a Heisenberg ferromagnet: The relevant symmetry of the Hamiltonian 
is the rotational symmetry in spin space, and the disordered and
ordered states are represented by the paramagnetic and the ferromagnetic
states, respectively. In the former, there is no preferred direction
for the magnetic moments of the spins. The net magnetization is
therefore zero, and the state has the same spin rotational symmetry
as the Hamiltonian. That is, if we rotate the coordinate system in
spin space, then the system will look the same. In the latter, on the other
hand, there {\it is} a preferred direction for the spins, and this is the
direction which the overall magnetization will point in. The state
therefore does no longer respect the spin rotational symmetry, and we
say that the symmetry is {\it spontaneously broken} (`spontaneously',
in order to distinguish this phenomenon from the explicit or external
breaking that occurs if we apply an external magnetic field). In an
ideal system the preferred direction is completely arbitrary, and in
real systems it is determined by very small external fields, like e.g.
the earth's magnetic field, or by a small breaking of the symmetry
in the system's Hamiltonian as provided by, e.g., the ionic lattice
strucuture. As a result, in an ideal system the magnetization in the
ordered state is not only non-zero, but it is also non-unique. Such a
thermodynamic quantitity that is zero in the disordered phase, and
non-zero and non-unique in the ordered phase, is called an {\it order
parameter}, a concept first introduced by Landau. The external field $B$
that couples to the order parameter, in our example the magnetic
field, is called the field {\it conjugate} to the order parameter.

As we approach the phase transition by changing some parameter, several
remarkable phenomena are observed. For instance, the correlations of the
order parameter become long-ranged. Let $M$ be the order parameter, and
let us consider the spatial correlations of $M$ (in the case of a magnet,
they can be measured by neutron scattering):

\begin{equation}
\langle M({\bf x})\,M({\bf y})\rangle = f(\vert {\bf x}-{\bf y}
    \vert/\xi)\quad.
\label{eq:1.1}
\end{equation}
Everywhere in parameter space except at the critical point, the
function $f$ decays in space on a length
scale $\xi$ called the {\it correlation length}. 
\footnote[2]{For most phase transitions, the function $f$ in
Eq.\ (\protect\ref{eq:1.1}) decays exponentially for large
arguments: $\ln f(x\rightarrow\infty) = -x + 0(\ln x)$. However,
in general the functional form can be more complicated.}
$\xi$ diverges
as the critical point is approached, usually like a power law that
is characterized by the correlation length critical exponent, $\nu$:
\begin{equation}
\xi \propto t^{-\nu}\quad,
\label{eq:1.2}
\end{equation}
where $t$ denotes some dimensionless distance in parameter space from 
the critical point. For instance, if the transition occurs at a
non-zero critical temperature $T_{\rm c}$, we can use 
$t=\vert T-T_{\rm c}\vert/T_{\rm c}$. At criticality, i.e. at
$t=0$, the correlation length diverges, which indicates that the
order parameter correlations decay only like a power law, which has
no intrinsic scale.
\footnote[3]{Mathematically, a power law is a
homogeneous function, while an exponential, for instance, is not.}
It is customary to denote this power law by an exponent
$\eta$:
\begin{equation}
\langle M({\bf x})\,M({\bf y})\rangle_{t=0} \propto \vert {\bf x} - {\bf y}
   \vert^{-d-2+\eta}\quad,
\label{eq:1.3}
\end{equation}
where $d$ denotes the spatial dimensionality of the system.

Apart from these long-range correlations in space, there are similar
effects in the temporal behavior of the system. Let us denote the
equilibration time, i.e. the time scale for the system to return to
equilibrium after it has been disturbed, by $\tau_c$. This equilibration
time diverges as criticality is approached, and it does so as a power
of the correlation length, with the power law characterized by an
exponent $z$:
\label{eqs:1.4}
\begin{equation}
\tau_c \propto \xi^z\quad,
\label{eq:1.4a}
\end{equation}
The inverse of $\tau_c$ defines a critical frequency scale $\omega_c$
that goes to zero as criticality is approached, a phenomenon called
{\it critical slowing down}:
\begin{equation}
\omega_c(t\rightarrow 0) \propto 1/\tau_c \rightarrow 0\quad.
\label{eq:1.4b}
\end{equation}

The exponents $\nu$, $\eta$, and $z$ that we have defined so far are
examples of {\it critical exponents}, which characterize the power law
behavior of various observables upon approach to the critical point.
Three other important critical exponents are $\beta$, which describes
the vanishing of the order parameter,
\label{eqs:1.5}
\begin{equation}
M(t\rightarrow 0) \propto t^{\beta}\quad,
\label{eq:1.5a}
\end{equation}
$\gamma$, which describes
the diverges of the order parameter susceptibility, which in our
example is the magnetic susceptibility $\chi = M/B$,
\begin{equation}
\chi(t\rightarrow 0) \propto t^{-\gamma}\quad,
\label{eq:1.5b}
\end{equation}
and the critical exponent $\delta$, which describes the dependence of
the order parameter on its conjugate field at criticality,
\begin{equation}
M(t=0,B\rightarrow 0) \propto B^{1/\delta}\quad,
\label{eq:1.6}
\end{equation}
Notice that a diverging susceptibility implies that there cannot be
a region where the order parameter depends linearly on the conjugate
field.

The set of critical exponents characterizes the critical behavior,
and it turns out that all of the critical exponents are not independent.
Rather, many of them are related to one another by {\it scaling laws}
or {\it exponent relations}. Furthermore, it turns out that the complete
set of critical exponents is the same for whole classes of phase
transitions. For instance, the critical exponents for the ferromagnetic
transitions in iron and nickel are the same, even though these
three metals have very different bandstructures, and their Curie
temperatures are very different. Perhaps even more surprisingly, the
critical exponents observed at the critical point of water are
the same as those at the liquid-gas critical points of the 
``quantum liquids'' He3 and He4. Phase transitions that 
share the same critical behavior, as expressed by the critical 
exponents, are said to belong to the same {\it universality class},
and the existence of such universality classes is referred to as
{\it universality} or {\it universal behavior}. It turns out that
universality classes are determined by basic symmetries of the
underlying Hamiltonian, and by the spatial dimensionality of the
system. The fact that the critical behavior is independent of the
microscopic details of the Hamiltonian is due to the diverging
correlation length: Close to a critical point, the system performs
an average over all length scales that are smaller than the (very
large) correlation length. This observation also gives a clue as to
how to theoretically deal with critical phenomena: In order to
correctly describe the universal critical behavior it should be
sufficient to work with an effective theory that keeps explicitly
only the asymptotic long-wavelength or large-distance behavior of the
original Hamiltonian. By contrast, observables
that vary from system to system even within a given unversality class,
like e.g. the critical temperature, are called {\it non-universal}
properties, and they are not accessible by means of such effective
theories.

The theoretical understanding of these, and other, properties of
phase transitions took place over a period of 100 years, starting
with van der Waals, continuing with Landau, and ending with Wilson.
Wilson's renormalization group, the essence of which is a set of scale 
transformations, provided us with a deep understanding of the
origin of universality and the critical power laws. More recently,
it has also turned out to be a very useful general tool for studying
the statistical mechanics of many-body systems in general, not only
close to some critical point. Here it is not our goal to describe the
renormalization group, for this purpose we refer the reader to the
many excellent expositions in the literature \cite{RG}. Rather, we will
take the critical phenomenology for granted, and on its basis discuss
the role of quantum mechanics in the context of phase transitions.
The only exception is Sec.\ \ref{subsec:II.3},
where we assume some familiarity with basic renormalization group
arguments in order to derive some results. As an explicit example, 
we will stick to ferromagnets, except in Sec.\ \ref{sec:III} which
deals with metal-insulator transitions.

\subsection{Classical versus quantum phase transitions}
\label{subsec:I.2}

A fundamental question is the following: To what extent is quantum
mechanics necessary in order to understand the critical phenomena
we have sketched in the previous subsection, and to what extent will
classical physics suffice? One can get the correct answer by means of
the following very simple considerations (which can be elaborated upon
if that is considered desirable). Generally speaking, quantum mechanics
is important whenever the temperature becomes lower than some characteristic
energy of the system under consideration. For instance, in an atom that
characteristic energy is the Rydberg energy. In our case, we have seen
that there is a characteristic frequency, namely $\omega_c$.
Let us assume that the corresponding energy
scale, $\hbar\omega_c$, is the smallest relevant energy scale. Since
$\omega_c(t\rightarrow 0)\rightarrow 0$, this is a reasonable guess close
to the transition. It then follows that quantum mechanics will be important
whenever the temperature $T$ obeys
\begin{equation}
k_{\rm B}T < \hbar\omega_c\quad.
\label{eq:1.7}
\end{equation}
Usually we think of phase transitions as taking place at some non-zero
critical temperature $T_c$, e.g. at the Curie temperature in our magnetic
example. It then follows that sufficiently close to the transition, namely
for
\begin{equation}
t < (T_c/T_0)^{1/\nu z}\quad,
\label{eq:1.8}
\end{equation}
we expect quantum mechanics not to be important for describing the system's
behavior. Here $T_0$ is some microscopic temperature scale, e.g., the Fermi
temperature in a metallic ferromagnet.
It follows that for any phase transition that takes place at a 
non-zero critical temperature $T_c > 0$, the critical behavior asymptotically 
close to the transition can be described entirely by classical physics.
This conclusion survives more rigorous arguments. These phase transitions
are called {\it classical} or {\it thermal} transitions. What drives the
correlation length to infinity are thermal fluctuation, which become very
large close to criticality.  In contrast, we
might think of a transition that occurs at zero temperature, and that is
triggered by varying some non-thermal parameter, e.g. the system's
composition. For instance, if $n$ is the concentration of some ingredient,
and $n_{\rm c}$ is the critical concentration, we can choose
$t = \vert n_{\rm c} - n\vert/n_{\rm c}$ as our dimensionless distance
from criticality. 
Then it obviously follows that quantum mechanics {\it will}
be important for describing the critical behavior. These transitions are
called {\it quantum} or {\it zero-temperature} phase transitions, and the
relevant fluctuatations are quantum fluctuations or 
zero-point motion. An example
would be a magnetic system that is, at $T=0$, continuously 
diluted with some non-magnetic
material until it undergoes a transition to the paramagnetic state. (We
will deal with the obviously relevant question of what happens at low,
but non-zero, temperatures in the following subsection.) Notice that,
according to this definition, some phase transitions in systems that are
usually considered quintessentially quantum mechanical, like the
superconducting transition in mercury at $T=4.2\,{\rm K}$, or the
$\lambda$-transition in helium at $T=2.17\,{\rm K}$, are classical
transitions. Indeed, in both cases the critical behavior (but {\it not}
the physics that triggers the transition, or the properties of either
phase) can be understood entirely by means of classical physics: Both
transitions are in the universality class of a classical 3-d xy-model.

\subsubsection{Classical phase transitions}
\label{subsubsec:I.2.A}

Let us now first consider some additional properties of classical
phase transitions. Within classical statistical mechanics, consider a
Hamiltonian
\label{eqs:1.9}
\begin{equation}
H(p,q) = K(p) + U(q)\quad,
\label{eq:1.9a}
\end{equation}
where $p$ and $q$ are the generalized momenta and positions, and
$K$ and $U$ are the kinetic and potential energy, respectively.
\footnote[4]{We exclude from our considerations systems of charged
particles in magnetic fields, and other cases of velocity
dependent potentials.}
The partition function
\begin{equation}
Z = \int dp\,dq\ e^{-H/k_{\rm B}T} = \int dp\ e^{-K/k_{\rm B}T} 
     \int dq\ e^{-U/k_{\rm B}T}\quad,
\label{eq:1.9b}
\end{equation}
then factorizes into a piece that depends only on $K$ and one that
depends only on $U$. As a result, one can study the system's static
properties independently from its dynamical ones. In particular,
the dynamical critical exponent $z$ is independent from all of the
other critical exponents, and the static critical behavior can be
studied, following Landau, by means of an effective functional of a 
time-independent order parameter. One often expresses this by saying
that `statics and dynamics decouple'.

Close to the critical point, the free energy density, 
$f = (-1/V)\,k_{\rm B}T\,\ln Z$, obeys a generalized homogeneity law,
\begin{equation}
f(t,B,\ldots) = b^{-d}\,f(t\,b^{1/\nu}, B\,b^{x_B},\ldots)\quad.
\label{eq:1.10}
\end{equation}
Here $V$ is the system volume, and $t$ and $B$ are the dimensionless
distance from the critical point and the external field conjugate to
the order parameter, respectively, as before. $b$ is an arbitrary
positive real number called a scale parameter, and Eq.\ (\ref{eq:1.10})
holds for all $b>0$. $\nu$ is the correlation length critical exponent,
and $x_B>0$ is a critical exponent that is related to $\delta$ by
$x_B = d\delta/(1+\delta)$. Since all thermodynamic quantities can be
obtained from the free energy, Eq.\ (\ref{eq:1.10}) provides us with
homogeneity laws for all of them. For instance, by differentiating $f$
with respect to $B$ we obtain a homogeneity law for the order parameter
density, $M = \partial f/\partial B$,
\begin{equation}
M(t,B) = b^{x_B - d}\,M(t\,b^{1/\nu}, B\,b^{x_B})\quad.
\label{eq:1.11}
\end{equation}
Since $b$ is arbitrary, we can in particular put $b=B^{-1/x_B}$. At
$t=0$ we then recover the above relation between $x_B$ and $\delta$.
By the same method, other exponent relations can be obtained, and it
turns out that of all the static critical exponents, only two are
independent, e.g. $\nu$ and $x_B$. Another useful substitution is to
set $b=t^{-\nu} \propto \xi$. This makes it obvious that letting
$b\rightarrow\infty$ is tantamount to approaching criticality. In this
context, a remark about the suppressed variables in Eq.\ (\ref{eq:1.10}),
which we denoted by ``$\ldots$'', is in order. They all enter 
Eq.\ (\ref{eq:1.10}) analogously to $t$ and $B$, but the exponents that
characterize their `scaled' entries on the right-hand side (i.e. the
analogs of $1/\nu$ and $x_B$), turn out to be negative. As a result, these
entries go to zero as one approaches criticality, and are called
`irrelevant variables' or `irrelevant operators'. If the observable under
consideration is a regular function of a particular variable for small 
values of the argument, then it follows that this variable becomes
unimportant as we approach criticality and does not influence the critical
behavior. It thus really becomes `irrelevant' in the ordinary sense of
the word. This is not the case, however, if the observable is a singular
function of some argument for small values of the argument. In this case
the irrelevant variable influences the critical behavior after all, and
one speaks of a `dangerous irrelevant variable'. Dangerous irrelevant
variables can cause substantial complications in the technical analysis
of scaling near critical points.

The homogeneity or scaling law, Eq.\ (\ref{eq:1.10}), was historically
first postulated phenomenologically, and it turned out that all observed
properties of phase transitions followed from it if appropriate values
for $\nu$ and $x_B$ were used, depending on the universality class under
consideration. It was the triumph of
Wilson's renormalization \cite{Wilson} group that
it allowed a derivation of the homogeneity law from first principles.
This derivation is highly technical, and we cannot go into it. Instead,
we refer the reader to the extensive literature on this subject \cite{RG}.

\subsubsection{Quantum phase transitions}
\label{subsubsec:I.2.B}

Let us now turn to the case of quantum phase transitions, i.e.
transitions that occur at $T=0$ and are triggered by some non-thermal
control parameter. As we have seen above, in this case quantum
mechanics is always important, and we need to employ quantum
statistical mechanics in order to calculate the partition function
and the free energy. Let the Hamilton operator of the system be
${\hat H}({\hat a}^{\dagger},{\hat a})$, with ${\hat a}^{\dagger}$
and ${\hat a}$ a collection of creation and annihilation operators.
In the usual imaginary time formalism, the time evolution of ${\hat a}$
is given by 
${\hat a}({\tau}) = e^{{\hat H}\tau}\,{\hat a}\,e^{-{\hat H}\tau}$,
with $\tau$ the imaginary time variable. A general theorem \cite{NegeleOrland}
then tells us that the partition function
for any quantum many-body system can be written as a function integral
of the form
\label{eqs:1.12}
\begin{equation}
Z = \int D[{\bar\psi},\psi]\ e^{S[{\bar\psi},\psi]}\quad.
\label{eq:1.12a}
\end{equation}
Here ${\bar\psi}$ and $\psi$ are space and imaginary time dependent
fields that are isomorphic to the sets of creation and annihilation
operators in a second quantization formulation of the problem. Their
nature depends on whether the quantum particles are fermions or bosons:
For the latter, the fields are classical or bosonic (i.e., they commute),
while for the former, they are fermionic (i.e., they anticommute). 
$D[{\bar\psi},{\psi}]$ is an appropriate integration measure defined
with respect to these fields \cite{FunctionalIntegration}. The {\it action}
$S[{\bar\psi},\psi]$ is uniquely determined by the Hamilton operator,
and is given by
\begin{eqnarray}
S[{\bar\psi},{\psi}]&=&\int d{\bf x} \int_{0}^{1/k_{\rm B}T} d\tau\ 
   {\bar\psi}({\bf x},\tau)\,\left[-\frac{\partial}{\partial\tau} +
   \mu\right]\,\psi({\bf x},\tau) 
\nonumber\\
&& - \int_{0}^{1/k_{\rm B}T} d\tau\ 
    H\left({\bar\psi}({\bf x},\tau),\psi({\bf x},\tau)\right)\quad.
\label{eq:1.12b}
\end{eqnarray}
Here $H$ as a function of ${\bar\psi}$ and $\psi$ has the same
functional form as ${\hat H}$ as a function of ${\hat a}^{\dagger}$ and
${\hat a}$.

Since the Hamiltonian taken at some imaginary time does not
commute with the Hamiltonian taken at another imaginary time,
we see that for quantum systems the statics and the dynamics are
intrinsically coupled and need to be treated together and
simultaneously. All phenomena that need to be described by means of
quantum statistical mechanics, and quantum phase transitions in
particular, therefore automatically fall under the title of this NATO ASI.
It further follows that for quantum phase transitions, in contrast to
classical ones, there are three independent critical exponents, and
the dynamical exponent $z$ needs to be determined together with the
static ones.

The homogeneity law for the free energy density, Eq.\ (\ref{eq:1.10}),
can now easily be generalized to the quantum case. From Eqs.\ (\ref{eq:1.2})
and (\ref{eq:1.4a},\ref{eq:1.4b}) 
we see that, as a function of $t$, frequencies scale
like $t^{\nu z}$. Furthermore, in the imaginary time formalism, temperature
and Matsubara frequencies are directly proportional to one another, and it
is therefore plausible that temperature and frequency will scale in the
same way. We thus add $T$ as an argument to our free energy density
and acknowledge the explicit $T$ in the definition of $f$ to obtain
\begin{equation}
f(t,T,B,\ldots) = b^{-(d+z)}\,f(t\,b^{1/\nu},T\,b^z,B\,b^{x_B},\ldots)\quad.
\label{eq:1.13}
\end{equation}
Comparing Eqs.\ (\ref{eq:1.10}) and (\ref{eq:1.13}) we see that a quantum
phase transition in $d$ spatial dimensions resembles the corresponding
classical transition in $d_{\rm eff} = d + z$ spatial dimensions! This
is also plausible from the point of view of the Landau functional (see
below), where a spatial integral $\int d{\bf x}$ in the classical case
gets replaced by a space-time integral $\int d{\bf x}\,d\tau$ in the
quantum case. Early work on the subject suggested that this observation
provides a fast and easy solution to the problem of quantum critical
behavior. However, as we will see, the argument is too superficial to
be reliable, and the extent to which it holds requires a careful and
detailed discussion.
\vfill\eject
\subsection{An example: The paramagnet-to-ferromagnet transition}
\label{subsec:I.3}

Let us illustrate the concepts introduced above by means of a concrete
example. For definiteness, we consider a metallic or itinerant 
ferromagnet.\footnote[5]{For the purposes of this subsection we might 
as well consider
localized spins, but the theory discussed in Sec.\ \ref{sec:II} below
applies to itinerant magnets only.}

\subsubsection{The phase diagram}
\label{subsubsec:I.3.A}

\begin{figure}[t]
\vskip 15mm
\centerline{\psfig{figure=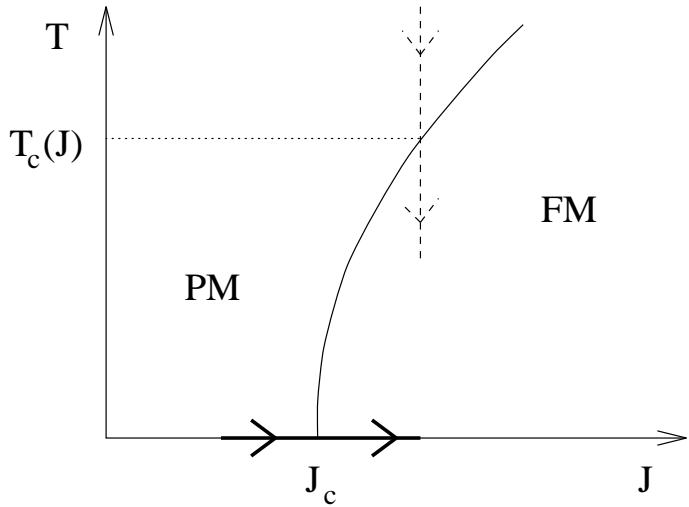,width=90mm}\vspace*{1mm}}
\vskip -10mm
\caption{Schematic phase diagram showing a paramagnetic (PM) and a
 ferromagnetic (FM) phase. The path indicated by the dashed line path
 represents a classical
 phase transition, and the one indicated by the solid line a
 quantum phase transition.}
\label{fig:1}
\end{figure}

Figure \ref{fig:1} shows a schematic phase diagram in the $T$-$J$ plane,
with $J$ the strength of the exchance coupling that is responsible for
ferromagnetism. The coexistence curve separates the paramagnetic phase
at large $T$ and small $J$ from the ferromagnetic one at small $T$ and
large $J$. For a given $J$, there is a critical temperature, the Curie
temperature $T_{\rm c}$, where the phase transition occurs. This is the
usual situation: A particular material has a given value of $J$, and the
classical transition is triggered by lowering the temperature through 
$T_{\rm c}$. Alternatively, however, we can image
changing $J$ at zero temperature (e.g. by alloying the magnet with some
non-magnetic material). Then we will encounter the paramagnet-to-ferromagnet
transition at the critical value $J_{\rm c}$. This is the quantum phase
transition we are interested in. Since we have seen that the quantum
transition is, loosely speaking, related to the classical one in a
different spatial dimensionality, and since we know that changing the
dimensionality usually means changing the universality class, we expect
the critical behavior at this quantum critical point to be different
from the one observed at any other point on the coexistence curve.

\begin{figure}[t]
\vskip 20mm
\centerline{\psfig{figure=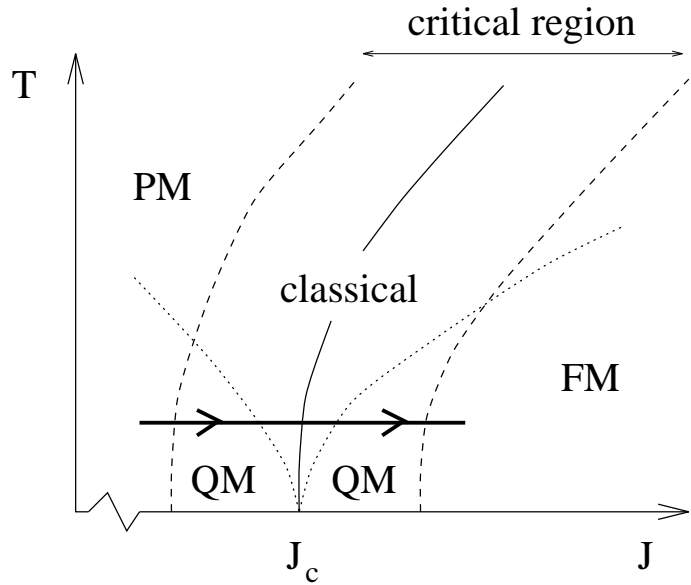,width=90mm}\vspace*{5mm}}
\vskip -20mm
\caption{Schematic phase diagram as in Fig.\ \protect\ref{fig:1} showing the
 vicinity of the quantum critical point $(J=J_{\rm c},T=0)$. Indicated are
 the critical region, and the regions dominated by the classical and quantum
 mechanical critical behavior, respectively. A measurement along the path
 shown will observe the crossover from quantum critical behavior away from
 the transition to classical critical behavior asymptotically close  to it.}
\label{fig:2}
\end{figure}

This brings us to the question of how continuity is guaranteed when one
moves along the coexistence curve. The answer, which was found by
Suzuki \cite{Suzuki}, also explains why the behavior at the $T=0$
critical point is relevant for observations at small but non-zero
temperatures. Consider Fig.\ \ref{fig:2}, which shows an enlarged
section of the phase diagram near the quantum critical point. The
critical region, i.e. the region in parameter space where the critical
power laws can be observed, is bounded by the two dashed lines. 
It then turns out that the critical region is divided into two
subregions, denoted by `QM' and `classical' in
Fig.\ \ref{fig:2}, in which the observed critical behavior is
predominantly quantum mechanical and classical, respectively.
The division between these two regions (shown as a dotted line in
Fig.\ \ref{fig:2}) is not sharp (and neither is the boundary of the
critical region), but rather a smooth {\it crossover} from
predominantly quantum mechanical to predominantly classical critical
behavior is observed as $J$ is increased at low but non-zero $T$.
The figure illustrates that the asymptotic critical behavior, very
close to the transition, is classical for all non-zero $T$, but that
for small $T$ there nevertheless is a sizeable region where quantum
critical behavior is observable. It also makes it clear that the
abovementioned continuity is realized by means of a distribution limit.

\subsubsection{Classic results for itinerant ferromagnets}
\label{subsubsec:I.3.B}

The subject of quantum magnetic phase transitions was pioneered by
Hertz \cite{Hertz}, who built on earlier work by Suzuki \cite{Suzuki}
and others \cite{BealMonod}. Hertz's main result can be recovered by
combining the above observation that the quantum phase transition
should correspond to the classical one in $d_{\rm eff} = d + z$
dimensions, with the fact that for classical magnets the upper
critical dimension $d_{\rm c}^+$, i.e. the dimensionality above which
the critical behavior is mean-field like, is $d_{\rm c}^+ = 4$. It then
follows that for the quantum transition, $d_{\rm c}^+ = 4-z$. Hertz
further found that $z=3$ for clean itinerant ferromagnets, and $z=4$
for disordered ones (we will reproduce his results in Sec.\ \ref{sec:II}
below). It then seemed to follow that $d_{\rm c}^+ = 1$ for clean
magnets, and $d_{\rm c}^+ = 0$ for disordered ones, so that the quantum 
critical behavior would be
mean-field like in all physical dimensions ($d=2,3$).

These classic results, which seemed to solve the problem (albeit the
answer was not very interesting from a theoretical point of view),
cannot be correct, however, since they violate other, very general
results that were obtained around the same time. Curiously, this
contradiction went unnoticed for more than 20 years. 

In 1974, Harris \cite{Harris} argued that at any critical point in a system 
that contains quenched disorder, the correlation length exponent must obey 
the inequality 
\begin{equation}
\nu \geq 2/d\quad,
\label{eq:1.14}
\end{equation} 
in order for the critical behavior to be stable with respect to the disorder. 
\footnote[6]{This is not how Harris formulated his criterion, but it
can be brought into this form by using suitable exponent relations.}
Harris' physical arguments were
later augmented by Chayes et al. \cite{Chayesetal}, who proved a
rigorous mathematical theorem to the same effect. Now the mean-field
value of $\nu$ is $\nu_{\rm MF} = 1/2$, which violates Harris' criterion
for all $d<4$. Hence Hertz's results for disordered systems cannot be
correct. It was also pointed out by Sachdev \cite{Sachdev_ZPhys} that Hertz's
results for clean systems in $d<1$ (an academic, but nonetheless 
interesting case) were at odds with some general scaling arguments.

It finally turned out that the classic results for both the clean and
the disordered case are indeed incorrect, for rather subtle and
interesting reasons. It also was shown that the theory can be salvaged
with relatively little effort, provided that one acknowledges the 
long-range correlations in itinerant electron systems that were explained
in the preceding lectures \cite{TRK's_lectures}.
This theory was developed in a series of
papers \cite{usMagnets}, the main arguments and results of which we 
reproduce in the next section.

\section{Quantum Critical Behavior of Itinerant Ferromagnets}
\label{sec:II}

We now sketch the derivation of an effective field theory for itinerant
ferromagnets, starting from a microscopic Hamiltonian. As we will see,
the derivation makes massive use of our knowledge of the properties of
{\it non-magnetic} electron systems. It is this knowledge that makes the
task of determining the correct quantum critical behavior much easier
than it would have been in the 1970s.

\subsection{Landau-Ginzburg-Wilson theory}
\label{subsec:II.1}

Let us consider a microscopic model for an itinerant ferromagnet. What
we have in mind is to develop an effective theory from such a miscroscopic
model that will be capable of correctly describing the quantum critical
behavior without being more detailed than necessary in other respects.
To this end we follow Landau's strategy \cite{Landau} of formulating a
theory in terms of an appropriate order parameter, and integrating out
all other degrees of freedom. For a ferromagnet, the relevant observable
is the spin density
\begin{equation}
{\bf n}_{\rm s}({\bf x},\tau) = {\bar\psi}({\bf x},\tau)\,{\vec\sigma}\,
   \psi({\bf x},\tau)\quad,
\label{eq:2.1}
\end{equation}
where the vector $\vec\sigma$ denotes the Pauli matrices. The relevant
electron-electron interaction is the spin-triplet interaction between the
spin density fluctuations,
\begin{equation}
S_{\rm int} = J \int d{\bf x} \int_{0}^{1/k_{\rm B}T} d\tau\ 
   {\bf n}_{\rm s}({\bf x},\tau)\,\cdot\,{\bf n}_{\rm s}({\bf x},\tau) 
    \quad.
\label{eq:2.2}
\end{equation}
The complete action is
\begin{equation}
S = S_{\rm 0} + S_{\rm int}\quad.
\label{eq:2.3}
\end{equation}
Here $S_{\rm 0}$ comprises all parts of the action other than $S_{\rm int}$
as defined in Eq.\ (\ref{eq:2.2}): It contains a term describing free
electrons (or lattice electrons\footnote[7]{For our purposes, which are
aimed at long-wavelength phenomena, it is irrelevant whether one considers
a lattice model or a continuum one.}) as well as electron-electron interactions
in all channels other than the spin-triplet one. In particular, $S_{\rm 0}$
contains the direct or Coulomb interaction between the electronic
charge densities. We will refer to the fictitious system that is described by
the action $S_{\rm 0}$ as the {\it reference ensemble}, and it will play an
important role in what follows.

Our goal is to calculate the partition function
\label{eqs:2.4}
\begin{equation}
Z = \int D[{\bar\psi},\psi]\ e^{S_{\rm 0}[{\bar\psi},\psi]}\,
    e^{J\int dx\ {\bf n}_{\rm s}(x)\cdot{\bf n}_{\rm s}(x)}\quad.
\label{eq:2.4a}
\end{equation}
Here and it what follows we adopt a four-vector notation 
$x\equiv ({\bf x},\tau)$
for space-time integrals. We now concentrate on the term $S_{\rm int}$, and
decouple it by means of a Gaussian or Hubbard-Stratonovich 
transformation \cite{HubbardStratonovich}. The latter consists of introducing
a classical (i.e. commuting) auxiliary field $M(x)$ that couples linearly to
the spin density, and rewriting $e^{S_{\rm int}}$ as a Gaussian integral,
\begin{equation}
Z = \int D[{\bar\psi},\psi]\ e^{S_{\rm 0}[{\bar\psi},\psi]}\,\int D[M]\ 
    e^{-J\int dx\ M^2(x) + 2J\int dx\ M(x)n_{\rm s}(x)}\quad.
\label{eq:2.4b}
\end{equation}
From now on, for simplicity we ignore the vector nature of the spin density.
Taking it into account just complicates the notation without leading to
qualitatively important effects. Next, we interchange the $M$ and $\psi$-integrations,
and formally carry out the latter,
\begin{equation}
Z = \int D[M]\ e^{-J\int dx\ M^2(x)}\ Z_{\rm 0}[M]\quad,
\label{eq:2.4c}
\end{equation}
where we recognize
\begin{equation}
Z_{\rm 0}[M] = \int D[{\bar\psi},\psi]\ e^{S_{\rm 0}[{\bar\psi},\psi] +
                2J\int dx\ M(x)\,n_{\rm s}(x)}\quad,
\label{eq:2.4d}
\end{equation}
as the partition function of the reference ensemble in an external `magnetic
field' that is proportional to $M(x)$. 
From the linear coupling between $M(x)$ and
the spin density it is clear that the expectation values of these two
quantities are proportional to one another. We thus identify $M(x)$ as the
order parameter field whose expectation value is the magnetization. It is
customary to define a Landau-Ginzburg-Wilson or LGW functional $\Phi[M]$,
\label{eqs:2.5}
\begin{equation}
\Phi[M] = J\int dx\ M^2(x) - \ln Z_{\rm 0}[M]\quad,
\label{eq:2.5a}
\end{equation}
where
\begin{equation}
Z_{\rm 0}[M] = \langle e^{2J\int dx\ M(x)\,n_{\rm s}(x)}\rangle_{\rm 0}
               \quad.
\label{eq:2.5b}
\end{equation}
Here $\langle\ldots\rangle_{\rm 0}$ denotes an average with respect to
the reference ensemble action $S_{\rm 0}$, and the partition function 
can be expressed in terms of the LGW functional as
\begin{equation}
Z = \int D[M]\ e^{-\Phi[M]}\quad.
\label{eq:2.5c}
\end{equation}

The above derivation of an order parameter theory for quantum ferromagets
is due to Hertz \cite{Hertz}. The salient point is that the LGW functional
is given in terms of the free energy of the reference ensemble in a
(space and time dependent) `external field' that is given by the order
parameter field. A more technical way to say this is that the LGW is given
by the generating functional for connected spin density correlation functions
in the reference ensemble, viz. $\ln Z_{\rm 0}$. Notice that all of the above
have been exact manipulations. The strategy underlying this exact rewriting
of the partition function is to express the free energy of the full system,
which undergoes a phase transition, in terms of that of the reference ensemble,
which does not.
\footnote[8]{Remember that the reference ensemble is missing the spin-triplet
part of the electron-electron interaction that is responsible for triggering
ferromagnetic phase transitions.}
The price one pays is that one needs to know the free energy of the latter in
the presence of an arbitrary space and time dependent magnetic field. As we
will see, however, enough is known about correlated electrons in magnetic
fields to successfully implement this strategy.

\subsection{The Landau expansion}
\label{subsec:II.2}

Follwing Landau, the next step is to expand the LGW functional in powers of $M$.
Remembering that the reference ensemble has no spontaneous magnetization, and
hence $\langle n_{\rm s}\rangle_0 = 0$, we have
\begin{equation}
Z_{\rm 0} = 1 + 2J^2 \int dx\,dy\ M(x)\,M(y)\,\langle n_{\rm s}(x)\,
                n_{\rm s}(y)\rangle_{\rm 0} + \ldots\quad,
\label{eq:2.6}
\end{equation}
where the terms not shown explicitly are of $O(M^4)$ or higher.
Taking a Fourier transform, we thus obtain for the LGW functional
\label{eqs:2.7}
\begin{eqnarray}
\Phi[M]&=&\frac{1}{V}\sum_{\bf q} T \sum_{\omega} M({\bf q},\omega)\,
          \left[1 - J\chi_{\rm s}(q,\omega)\right]\,M(-{\bf q},\omega)
\nonumber\\
&&\qquad + \sum_{\{{\bf q},\omega\}} u_4(\{{\bf q},\omega\})\,M^4({\bf q},
                  \omega) + \ldots\quad,
\label{eq:2.7a}
\end{eqnarray}
where $q = \vert{\bf q}\vert$, and
\begin{equation}
\chi_{\rm s}(q,\omega) = \langle n_{\rm s}({\bf q},\omega)\,
                           n_{\rm s}(-{\bf q},-\omega)\rangle_0^{\rm c}\quad,
\label{eq:2.7b}
\end{equation}
is the connected two-point spin-density correlation function of the reference
ensemble, i.e., its spin susceptibility,
\begin{equation}
u_4 = \langle n_{\rm s}\,n_{\rm s}\,n_{\rm s}\,n_{\rm s}\rangle_0^{\rm c}\quad,
\label{eq:2.7c}
\end{equation}
is the corresponding connected four-point function, etc. We have used an
obvious schematic notation for the quartic terms, since we will not study
them in detail here.

The salient point of this formal development is that the LGW functional is
given in terms of the connected spin density correlation functions of the
reference ensemble. The reference ensemble, however, is a Fermi liquid,
or its generalization to the case of quenched disorder, and
so its correlation functions are known! Indeed, the 
preceding lectures \cite{TRK's_lectures} has
explained in some detail what they are, and we can now draw on that knowledge.
For the sake of definiteness we will discuss only disordered systems,
\footnote[9]{Strictly speaking, we should have used a replicated theory
to deal with the quenched disorder. In the interest of keeping our pedagogical
discussion simple we have suppressed this technical point. For an introductory
discussion of the replica trick, see Ref.\ \protect\cite{Grinstein}.}
where the effects we want to demonstrate are most pronounced. Qualitatively
similar, albeit weaker, phenomena are present in clean systems as well, as
has been discusssed in the original literature \cite{usMagnets}.
The appropriate limit to study the correlation functions is that of small
wavenumbers and frequencies, $q,\omega \rightarrow 0$, with $\omega\ll q$
\footnote[10]{Otherwise one does not reach criticality, which can be seen as 
follows. Since the magnetization is conserved,
ordering on a length scale $L$ requires some spin density to be transported
over that length, which takes a time $t\propto L^2/D$, with $D$ the spin
diffusion coefficient. Now look at the system at a momentum scale $q$ or a
length scale $L\propto 1/q < \xi$, with $\xi$ the coherence length. Because of
the time it takes the system to order on that scale, the condition for
criticality is $L^2 < {\rm Min}(Dt,\xi^2)$. In particular, one must have
$L^2 < Dt$, or $\omega \propto 1/t < Dq^2$.}
in suitable units.
\footnote[11]{For instance, one can measure $q$ in units of the Fermi wavenumber
$k_{\rm F}$, and $\omega$ in units of $Dk_{\rm F}^2$, with $D$ the spin
diffusion coefficient of the reference ensemble.}
As we have seen in the preceding lectures \cite{TRK's_lectures,chi_s}
the spin susceptibility of a
disordered Fermi liquid in this limit reads
\label{eqs:2.8}
\begin{equation}
\chi_{\rm s}(q,\omega) = \chi_{\rm s}^0(q)\,\left[1 - \vert\omega\vert/Dq^2
                          + \ldots\right]\quad,
\label{eq:2.8a}
\end{equation}
and the long-wavelength expansion of the static spin susceptibility 
$\chi_{\rm s}^0$ reads
\begin{equation}
\chi_{\rm s}^0(q) = {\rm const.} - q^{d-2} - q^2 + o(q^2)\quad,
\label{eq:2.8b}
\end{equation}
where $o(q^2)$ denotes terms that are smaller than $q^2$, and
we have omitted positive prefactors of all terms in the expansion, as
they will not be important for what follows. The LGW functional now takes
the form
\begin{eqnarray}
\Phi[M]&=&\frac{1}{V}\sum_{\bf q} T\sum_{\omega} M({\bf q},\omega)\,
          \left[t + q^{d-2} + q^2 + \vert\omega\vert/q^2\right]\,
          M(-{\bf q},-\omega) 
\nonumber\\
&&\qquad\qquad\qquad\qquad\qquad + O(M^4)\quad.
\label{eq:2.9}
\end{eqnarray}
The four-point correlation function $u_4$ also contains nonanalyticities,
which show up the in the quartic term in $\Phi[M]$, and the same is
true for all higher order terms. While a careful study of these terms is
necessary for a complete treatment of the problem, we suppress them
here for brevity and simplicity, and refer the interested reader to
the original literature where they have been discussed in 
detail \cite{usMagnets}.

Before we analyze the LGW functional,
Eq.\ (\ref{eq:2.9}), in the next subsection, two
remarks are in order. First, if we had used a reference ensemble of
non-interacting electrons, rather than our more realistic one that
includes electron-electron interactions, then we would have missed
the non-analytic terms proportional to $q^{d-2}$ in Eqs.\ (\ref{eq:2.8b}) and
(\ref{eq:2.9}). As a consequence, we would have recovered Hertz's
results which, as we have seen earlier, cannot be correct. While it
seems at this point as if a realistic choice of the reference ensemble
were crucial, this is disturbing from some fundamental theoretical
points of view. Indeed, a careful investigation reveals that the
precise choice of the reference ensemble is {\it not} important.
We will come back to this point in Sec.\ \ref{subsec:II.4} below.
Second, the physical origins of the $q^{d-2}$ term are long-range spatial
correlations in the reference ensemble of interacting electrons that
underlies our effective LGW action, as has been explained in the
preceding lectures \cite{TRK's_lectures}.
These long-range spatial correlations are in turn
a consequence of the coupling of statics and dynamics in quantum
systems. We now see what has happened: In addition to the order parameter
fluctuations that develop a long range near criticality, our electron 
systems also contains long-range correlated degrees of freedom (viz.
the `diffusons' of the preceding lectures) that have nothing to do 
with phase transition physics, and that are present even far away from
the transition. These degrees of freedom have been integrated out in
our derivation of the LGW functional, which by definition is a functional
of the order parameter field only. The inevitable consequence of this
integrating out of slow modes is nonanalyticities in the resulting
LGW functional. Such nonanlyticities violate the very spirit of the
LGW concept, and to perform explicit calculations for the
non-local field theory given by Eq.\ (\ref{eq:2.9}) would be very
difficult indeed. An obvious way to avoid these problems would be to
{\it not} integrate out the diffusons, but rather derive a generalized
LGW theory in terms of {\it all} the soft modes in the systems.
However, as we will see in the next subsection, for
the purpose of determining the critical behavior the non-local theory
can be handled, and the present route is the fastest one to answer the
questions we have asked.

\subsection{Renormalization group analysis}
\label{subsec:II.3}

To finish our treatment of itinerant ferromagnets, we now analyze the
LGW functional, Eq.\ (\ref{eq:2.9}), by means of power counting or a
tree-level renormalization group (RG) analysis. Space and time constraints
do not allow us to explain this technique here. Readers not familiar with
it can find excellent and very accessible treatments in
Refs.\ \cite{ClassicalPhaseTransitions}.

Let us restrict ourselves to spatial dimensions $d < 4$, where the
$q^{d-2}$ term dominates over the analytic $q^2$ term, and let
us look for a Gaussian RG fixed point where neither the $q^{d-2}$
term nor the $\vert\omega\vert/q^2$ term in Eq.\ (\ref{eq:2.9}) are
renormalized. At such a fixed point, the critical order parameter correlation
function will behave like
\label{eqs:2.10}
\begin{equation}
\langle M\,M\rangle_{t=0} \propto 1/q^{d-2} \equiv q^{2-\eta}\quad,
\nonumber
\end{equation}
where the last relation reflects the definition of the critical exponent
$\eta$, see Eq.\ (\ref{eq:1.3}). We thus have
\begin{equation}
\eta = 4-d\quad.
\label{eq:2.10a}
\end{equation}
Furthermore, at such a Gaussian fixed point the frequency clearly scales with
the wavenumber like $\omega \sim q^d \equiv q^z$,
\footnote[12]{We use $\propto$ for `proportional to', and $\sim$ for `scales
like' or `has the same scale dimension as'.}
which yield a dynamical
critical exponent
\begin{equation}
z = d\quad.
\label{eq:2.10b}
\end{equation}
Finally, the exponent $\gamma$ for the Gaussian fixed point can also just be
read off the Gaussian action: At zero frequency and wavenumber, we have
\begin{equation}
\langle M\,M\rangle_{q=\omega=0} \propto 1/t \equiv t^{-\gamma}\quad,
\nonumber
\end{equation}
so that we have
\begin{equation}
\gamma = 1\quad.
\label{eq:2.10c}
\end{equation}
In order to determine the correlation length exponent $\nu$, we define the
scale dimension $[q]$ of a wavenumber $q$ to be $[q]=1$. The requirement
that the $q^{d-2}$ term in the LGW functional be dimensionless then leads
to a scale dimension of the order parameter field of $[M] = -(d-2)/2$.
Since the $t$ term must also be dimensionless, this yields
\begin{equation}
t \sim q^{d-2} \sim \xi^{-(d-2)} \equiv \xi^{-1/\nu}\quad,
\nonumber
\end{equation}
from which we read off $\nu$ as
\begin{equation}
\nu = 1/(d-2)\quad.
\label{eq:2.10d}
\end{equation}
All of these results obviously hold only for $2<d<4$. For $d\leq 2$ the
electrons in the reference ensemble become localized (see the preceding
lectures) and our theoretical framework breaks down, while for $d>4$ the
$q^2$ term dominates over the $q^{d-2}$ term, and we recover
Hertz's mean-field exponents, i.e. $\eta=0$, $z=4$, $\gamma=1$, and 
$\nu=1/2$.

A quick check shows that our result fulfils the Harris criterion, 
$\nu\geq 2/d$ (see Sec.\ \ref{subsubsec:I.3.B} above) for all
values of $d$. While this is encouraging, it is of course only
a necessary criterion for our results representing the correct
quantum critical behavior, not a sufficient one. To establish
the latter, one needs to consider the higher order terms in the LGW
functional, and establish that our Gaussian fixed point is stable.
At tree level, this has been done in the original 
literature \cite{usMagnets}, and the
result was that the fixed point is indeed stable. The analysis of the
quartic term in $\Phi$ in particular also yields the equation of state,
and thus the critical behavior of the order parameter itself, i.e. the
critical exponents $\beta$ and $\delta$. The result is
\begin{equation}
\beta = 2/(d-2)\qquad,\qquad\delta = d/2\quad,
\label{eq:2.10e}
\end{equation}
for $2<d<6$, while for all $d>6$ these exponents have their mean-field
values $\beta=1/2$ and $\delta=3$, respectively. The analysis of the higher
order terms thus establishes $d=6$ as another upper critical dimensionality,
in addition to $d=4$. In $d=4$ and in $d=6$ logarithmic corrections to
scaling occur, as is usually the case at an upper critical dimensionality.

Finally, we mention again that an analogous analysis of clean systems
yields qualitatively very similar results. Essentially, the exponent
$d-2$ in the non-analytic terms gets replaced by $d-1$, which leads to a
single upper critical dimension $d_{\rm c}^+ = 3$. Contrary to the
disordered case, the critical behavior in the most interesting dimension
$d=3$ is therefore mean-field like with logarithmic corrections to
scaling \cite{usMagnets}.

\subsection{Final remarks}
\label{subsec:II.4}

We finish this section with a few additional remarks.
\footnote[13]{Like the preceding subsection,
 some of these remarks require familiarity with renormalization group
 techniques.}
As we have sketched in the preceding subsection, and as has been more
carefully established in the original literature, we have managed to
determine the critical behavior at the quantum ferromagnetic transition
exactly, yet this behavior is not mean-field like. This is surprising,
as usually non-mean field like critical behavior cannot be obtained
exactly, save for a very few models. To find out what has enabled
us to do so, let us look again at our LGW functional, Eq.\ (\ref{eq:2.9}).
One way to state the effect of the non-critical soft modes that have led
to the $q^{d-2}$ term is to say that they have established an effective
long-range interaction between the order parameter fluctuations. Indeed,
a Fourier transform of the non-analytic term yields an interaction that
falls off like $1/r^{2(d-1)}$. It is well known that such long-range
interactions stabilize Gaussian critical behavior that is not mean-field
like \cite{FisherMaNickel}. What is remarkable here is that this long-range
interaction is not put in by hand, but rather is generated by the system
itself via the non-critical slow modes.

These considerations raise the question of whether the phenomenon
discussed here is germane to quantum phase transitions. In principle, it
is not. Whenever there are slow modes in addition to the order parameter
fluctuations that couple to the latter, one will obtain nonanalyticities
in the LGW functional, and resulting unusual critical behavior,
irrespective of whether one deals with a classical or a quantum phase 
transition. However, quantum transitions are much more susceptible to this
mechanism, for the simple reason that there are many modes that are soft
only at $T=0$ and acquire a mass at non-zero temperature. Our diffusons are
a good example: As was discussed in the preceding lectures, they are indeed
massive at $T>0$.

Finally, we come back to a point raised at the end of Sec.\ \ref{subsec:II.2}.
There we mentioned that whether or not one obtains the correct
critical behavior seems to depend 
crucially on the choice of the reference ensemble.
While one might argue that a non-interacting reference ensemble is simple
not a realistic model, this leads to the following paradox: Suppose we
consider a model whose action consists only of a free electron part, and
the spin-triplet interaction $S_{\rm int}$ of Eq.\ (\ref{eq:2.2}). Then
a decoupling of $S_{\rm int}$, as performed above, seems to lead to mean-field
critical behavior. However, if we considered some fraction of $S_{\rm int}$
a part of $S_0$, and decoupled the rest, then we would have an interacting
reference ensemble and would obtain the above non-mean field like critical
behavior. Clearly, both procedures are equally valid and should lead to the
same result. The resolution of this paradox is as follows. In the case of an
interacting reference ensemble, already the bare LGW functional contains the
crucial nonanalyticities. Therefore, a RG analysis at tree level is sufficient
to obtain the correct critical behavior. In the case of a non-interacting
reference ensemble, on the other hand, the bare action does not contain the
crucial non-analytic terms, but they are generated if the RG analysis is
carried to higher order in the loop expansion. Indeed, an inspection of
Hertz's model shows that the $q^{d-2}$ term is indeed generated by the RG,
starting at one-loop order. Since the generated term is relevant with 
respect to the mean-field fixed point, it invalidates the zero-loop
analysis. These observations serve as a reminder of a fact that is 
well-known in principle, but occasionally forgotten: Any RG analysis to a
given order in a loop expansion gives the correct answer only if no
relevant new terms in the action are generated at higher order. 
Unfortunately, {\it proving} that no such terms are generated at any
order in the loop expansion amounts to
proving that the theory is renormalizable, a task that is very difficult
and has been done only for a few select models.

\section{The Anderson-Mott Transition}
\label{sec:III}

At the end of these lectures, we would like to briefly touch upon some
very different quantum phase transitions, namely the types of metal-insulator
transition known as Anderson, and Anderson-Mott transitions, respectively. 
Our motivation for doing so
is chiefly to dispel any possible misconception that the concepts developed
so far apply only to quantum phase transitions that are magnetic in nature,
or that are represented as the $T=0$ end point of a line of classical phase
transitions. Indeed, the Anderson-Mott transition is neither. We would also
like to illustrate the point that the scaling ideas that have been so 
successful in connection with phase transitions, classical and quantum 
mechanical, may be applied to transport coefficients 
as well as to thermodynamic quantities \cite{HohenbergHalperin}.
Apart from making
these two point, the time and space we devote to this subject are grossly 
inadequate. Extensive reviews can be found in
Refs.\ \cite{R,usMarch,LeeRamakrishnan,MacKinnonKramer}.

\subsection{The simplest metal-insulator transition: The classical Lorentz 
                                                                       model}
\label{subsec:III.1}

\begin{figure}[ht]
\centerline{\psfig{figure=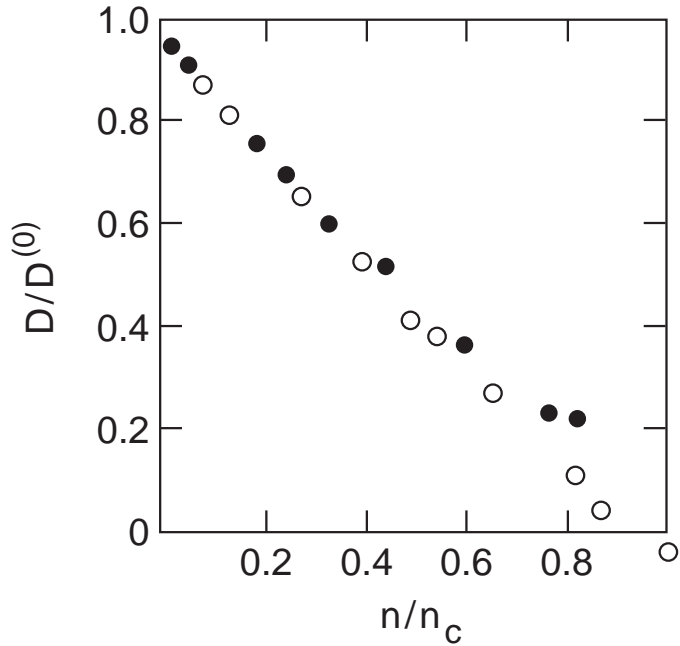,width=90mm}\vspace*{5mm}}
\vskip 0mm
\caption{Numerical simulation data for the diffusion coefficient $D$ versus
 the scatterer density $n$ of a $2$-$d$ classical Lorentz model. Full symbols
 represent data by Bruin \protect\cite{Bruin}, and open ones data by
 Alder and Alley\protect\cite{AlderAlley}, and by
 Alley\protect\cite{Alley}.
 $D$ is normalized by its Boltzmann value $D^{(0)}$, and $n$ by its
 critical value $n_{\rm c}$. From Ref.\ \protect\cite{R}.}
\label{fig:3}
\end{figure}

The simplest metal-insulator transition occurs in a classical model, viz.
the classical Lorentz gas. It is well known that with increasing scatterer
density $n$, the diffusivity $D$ of the moving particle decreases (see
the preceding lectures), until
if finally reaches zero at a critical scatterer density $n_{\rm c}$.
This behavior can be seen in the numerical data shown in Fig.\ \ref{fig:3}.
With $t=\vert n - n_{\rm c}\vert/n_{\rm c}$ the dimensionless distance
from the critical point, one has
\begin{equation}
D(t\rightarrow 0) \propto t^s\quad,
\label{eq:3.1}
\end{equation}
with $s$ a critical exponent. The reason for the localization of the
diffusing particle is that with increasing scatterer density it is
increasingly likely to get trapped in cages of scatterers from which
it cannot escape.

\subsection{Electrons: The Anderson transition, and the Mott transition}
\label{subsec:III.2}

Let us now add quantum mechanics to these considerations, and consider the
quantum Lorentz model that was discussed in detail in the preceding lectures.
Quantum mechanics has two competing effects on the transport properties:
On the one hand, it should increase the diffusivity, since it allows the
particle to tunnel out of cages it would be trapped in classically. On the
other hand, it leads to the quantum interference or weak-localization 
effects that enhance the backscattering amplitude. As we saw in the
preceding lectures, the latter effect wins. The quantum Lorentz model
therefore has a stronger tendency to localize the particle than the 
classical one,
to the point that in $d=2$ the particle is localized even for arbitrarily
small scatterer density. In $d>2$, however, there is a metal-insulator
transition at a finite value of $n$, which is called an Anderson
transition.

The Anderson transition is a strange phase transition from a statistical
mechanics point of view, as it has no simple order parameter, and no
upper critical dimensionality. However, it has a lower critical dimensionality,
$d_{\rm c}^- = 2$, as for $d\leq 2$ there is no transition. This has been
exploited for studies of the Anderson transition in $\epsilon = d-2$
expansions. More generally, the dynamical conductivity $\sigma$ 
(or, alternatively,
the diffusion coefficient $D$) obeys a generalized homogeneity 
law \cite{Wegner76}
\begin{equation}
\sigma (t,\omega) = b^{-(d-2)}\,\sigma (t\,b^{1/\nu},\omega\,b^z)\quad.
\label{eq:3.2}
\end{equation}
Putting $\omega=0$, and $b=t^{-\nu}$, we obtain from Eq.\ (\ref{eq:3.2})
the behavior of the static conductivity,
\label{eqs:3.3}
\begin{equation}
\sigma(t,0) \propto t^s\quad,
\label{eq:3.3a}
\end{equation}
with a conductivity exponent
\begin{equation}
s = \nu (d-2) \geq 2(d-2)/d\quad.
\label{eq:3.3b}
\end{equation}
The equality in Eq.\ (\ref{eq:3.3b}) is known as Wegner's scaling law, and
the inequality results from the Harris criterion, Eq.\ (\ref{eq:1.14}).
Due to the poor convergence properties of the $2+\epsilon$ expansions,
no reliable theoretical values for $s$ or $\nu$ are available. Numerical
calculations in $d=3$ yield values for $\nu$ in the range 
1.3 - 1.5 \cite{MacKinnonKramer}.

The quantum Lorentz model does not contain any electron-electron interaction,
and the Anderson transition is therefore entirely driven by disorder. The
opposite case, namely a metal-insulator transition that is entirely driven
by interactions, with no quenched disorder present, was proposed
by Mott to explain why certain materials with one electron per unit cell,
e.g. NiO, are insulators. Mott's original idea hinged on the long-range
nature of the Coulomb interaction, and in this case the metal-insulator 
transition comes about by means of a breakdown of screening and is of 
first order \cite{Mott}. A similar,
albeit continuous, transition is believed to occur in a model with a 
short-ranged electron-electron interaction known as the Hubbard model.
This Mott-Hubbard transition is still not well understood in $d=3$, although
much progress has been made recently on high-dimensional 
models \cite{HubbardReview}.

\subsection{The Anderson-Mott transition}
\label{subsec:III.3}

\begin{figure}[t]
\centerline{\psfig{figure=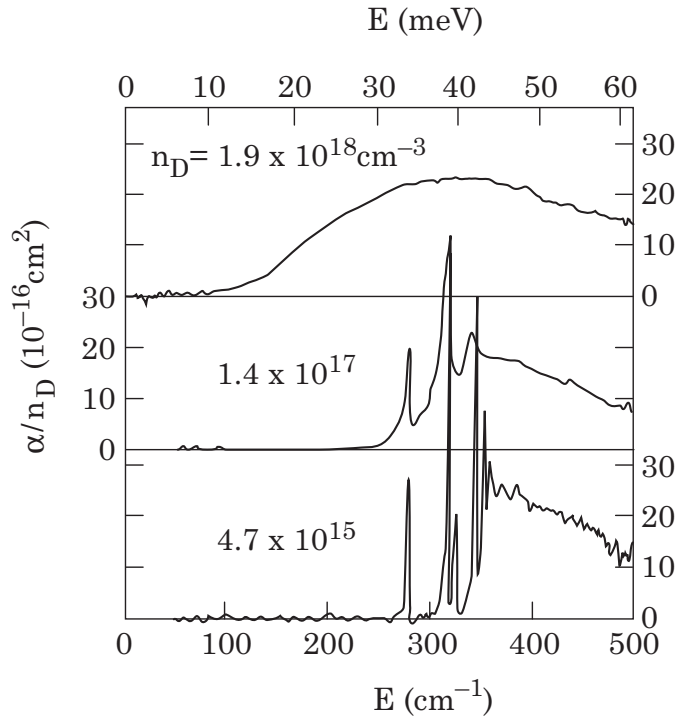,width=90mm}\vspace*{5mm}}
\vskip 0mm
\caption{Infrared absorption coefficient $\alpha$ for three different
 donor concentrations $n_{\rm D}$ in Si:P as measured by 
 Thomas at al. \protect\cite{Thomasetal}.
 The critical concentration in
 this system is $n_{\rm c}\approx 3.7\times 10^{18}\,{\rm cm}^{-3}$.
 From Ref.\ \protect\cite{R}.}
\label{fig:4}
\end{figure}

\begin{figure}[t]
\centerline{\psfig{figure=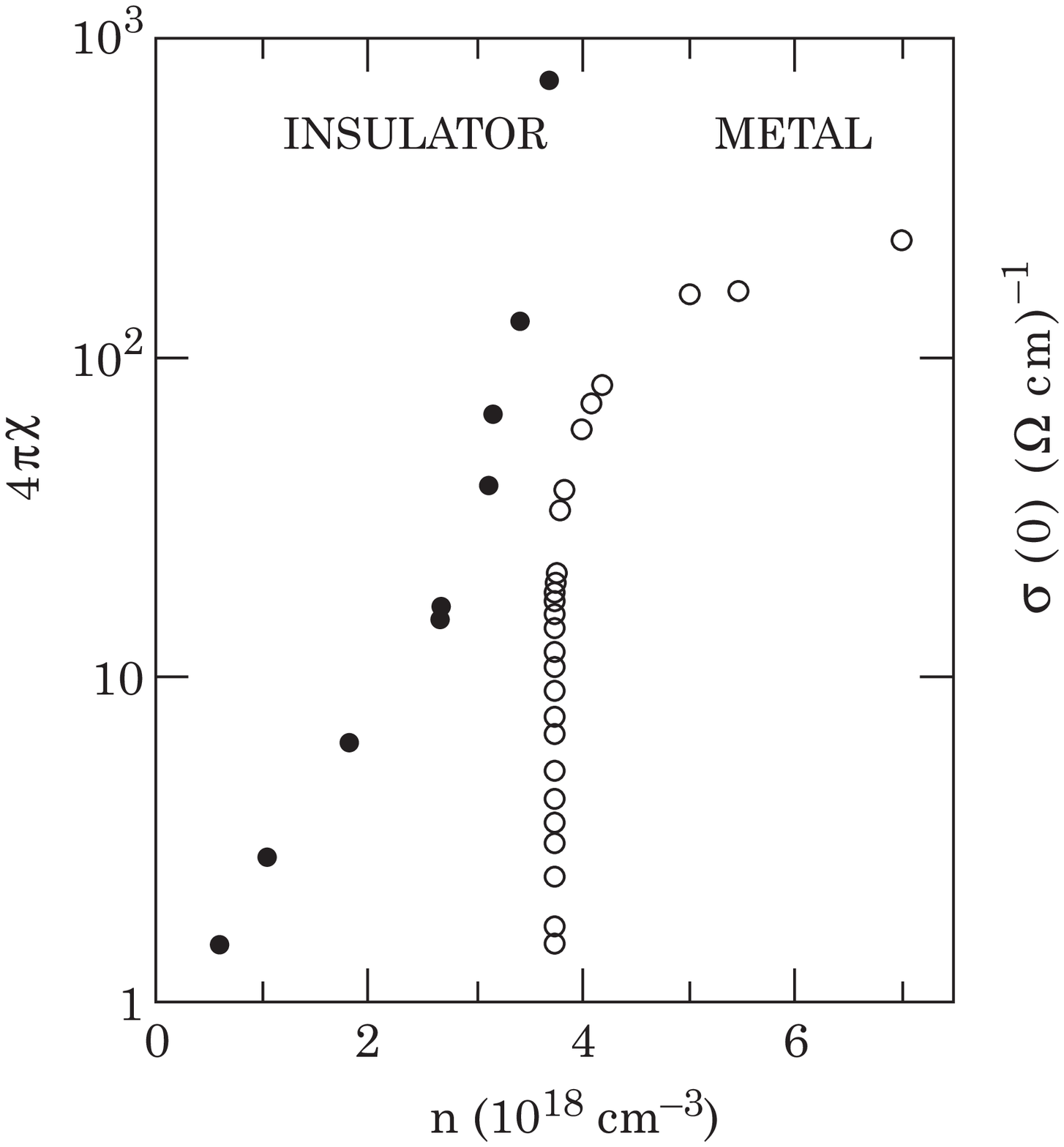,width=90mm}\vspace*{5mm}}
\vskip 0mm
\caption{Divergence of the dielectric susceptibility $\chi$ (full symbols),
 and vanishing of the static conductivity $\sigma$ (open symbols), both
 extrapolated to zero temperature, at the metal-insulator transition in
 Si:P as measured by Rosenbaum et al. \protect\cite{Rosenbaumetal}.
 From Ref.\ \protect\cite{R}.}
\label{fig:5}
\end{figure}

An Anderson-Mott transition results if one combines the driving forces for
the Anderson and Mott transitions and considers the case of interacting
electrons in the presence of disorder. This is a problem of great
experimental interest. For instance, the metal-insulator transition that
is observed in doped semiconductors as a function of the dopant
concentration is believed to be of that type. Fig.\ \ref{fig:4}
shows the infrared absorption spectrum of phosphorus-doped silicon.
For very low dopant concentrations one sees a hydrogen-like spectrum that
is produced by isolated phosphorus atoms. With increasing dopant concentration
these `atoms' start to overlap, which leads to a broadening of the spectral
features. At the highest concentration shown in the figure, the spectrum is
smooth, but it still represents an insulator (no absorption at zero
energy). With further increasing donor concentration, the system undergoes
a quantum phase transition to a metal, as can be seen in Fig.\ \ref{fig:5},
where both the static conductivity and the dielectric susceptibility are
plotted versus the phosphorus concentration. This system, and some other
doped semiconductors, are the experimentally best studied examples of
metal-insulator transitions. Notice that this is a quantum phase transition
that has no classical counterpart, as there can be no true insulator at
any non-zero temperature.

Much theoretical effort has been devoted to the Anderson-Mott transition.
These approaches fall into two distinct classes. The first one contains
$\epsilon$-expansions about the lower critical dimension 
$d_{\rm c}^-=2$ \cite{R}.
These theories lead again to Wegner scaling, Eq.\ (\ref{eq:3.2}), as in
the case of the Anderson transition, and Wegner's scaling law, 
Eq.\ (\ref{eq:3.3b}), still holds. This leads to $s=\nu\geq 2/3$ in
$d=3$, which contradicts the experimental result that the value of $s$ is
close to $1/2$.
\footnote[14]{All experimentalist do not agree with this result. 
Ref.\ \cite{vonLoehneysen} has reported $s>1$ for Si:P.}

The second class is formed by an order parameter 
theory \cite{usMarch} that uses
the fact that the density of states at the Fermi level may be considered
an order parameter for the Anderson-Mott transition (in contrast to the
case of the Anderson transition, where the density of states is uncritical.)
This theory has established $d_{\rm c}^+=6$ as the upper critical dimension
for the Anderson-Mott transition (again in contrast to the Anderson
transition, where $d_{\rm c}^+=\infty$), and the critical behavior for
$d>6$ is exactly known and mean-field like. Certain technical
analogies between this theory and theories for magnets in random magnetic
fields have led to the suggestion that in $d<6$, and in particular in
$d=3$, the Anderson-Mott transition has aspects that are reminiscient of
a glass transition, with exponential rather than power-law critical behavior
for many observables. A scaling theory has been developed for this
unorthodox critical behavior \cite{usMarch}, 
which is at least consistent with existing
experimental results, but so far no microscopic theory exists.

\acknowledgements
We would like to thank our collaborators on some of the work on ferromagnetic
systems reviewed above, Thomas Vojta and Rajesh Narayanan.
This work was supported by the NSF under grant numbers 
DMR--96--32978 and DMR--98--70597. This research was supported in part
by the National Science Foundation under grant No. PHY94-07194.

\vfill\eject
\end{document}